\documentclass[12pt]{article}
\usepackage[dvips]{graphics}

\begin{document}
\begin{center}
\LARGE {The spin of the mesons and baryons}
\bigskip

\Large {E.L. Koschmieder}
\medskip

\small {Center for Statistical Mechanics\\
The University of Texas at Austin, Austin TX 78712\\
e-mail: koschmieder@mail.utexas.edu}\\
\medskip
\small {February 03, 2003}
\end{center}

\bigskip

\noindent
\small
{It is shown that the spin of the $\pi^0, \eta, \Lambda, \Sigma^{\pm,0}, 
\Xi^{-,0}, \Lambda^+_c, \Sigma_c^{0},\, \Xi_c^{0},$ and $\Omega_c^0$ mesons and 
baryons can be explained by the sum of the angular momentum vectors and the 
spin vectors of the electromagnetic waves which are in these particles  
according to the standing wave model. The spin of the  $\pi^\pm$, K$^{\pm,0}$,  
D$^{\pm,0}$, and D$_S^\pm$ mesons and of the neutron is the sum of the angular 
momentum vectors of the oscillations and of the spin vectors of the neutrinos 
and the electric charges which are in the 
cubic lattice of these particles. Spin 1/2 is the consequence of the 
superposition of two perpendicular standing waves of equal frequencies and 
amplitudes shifted in phase by $\pi$/2. The spin of the antiparticles of the mesons and baryons is the same as the spin of the ordinary particles.} \normalsize

\section*{Introduction}

The spin or the intrinsic angular momentum is, after the mass, the 
second most important property of the elementary particles. As is well-known the spin 
of the electron was discovered by Uhlenbeck and Goudsmit [1] more than 75 
years ago. Later on it was established that the baryons have spin as well, 
but not the mesons. An explanation of the spin of the particles has sofar 
not been found.  Biedenharn and Louck [2] state on p.\,5 that, contrary to 
widespread perceptions, ``internal angular momentum is a classical, 
nonrelativistic concept" and that ``spin\,1/2 internal angular momentum is also 
a classical nonrelativistic field-theoretic concept". The concept that 
spin 1/2 is a relativistic effect originates from Dirac's equation. 
However, since electrons with practically zero 
velocity still have spin it is hard to see how spin 1/2 can be a 
relativistic effect. It is 
also difficult to see how the spin of a particle can be 
explained without consideration of the structure of the particle.  
Actually it appears to be 
crucial for the validity of a model of the elementary particles that the 
model can also explain the spin of the particles without additional 
assumptions. The standard model of the particles cannot explain the 
spin because the spin is imposed on the quarks. For current efforts to 
understand the spin of the nucleon see Jaffe [3] and of the 
spin structure of the $\Lambda$ baryon see G\"ockeler et al. [4]. 

We will look at the spin as a property of the standing waves in the 
cubic lattice of the particles which we have described in [5]. We 
cannot repeat the points made there. May it suffice to mention that we 
showed that the spectrum of the so-called stable mesons and baryons 
consist of a gamma-branch with the $\pi^0, \eta, \Lambda, \Sigma^0, 
\Xi^0, \Omega^-, \Lambda^+_c, \Sigma_c^0,\, \Xi_c^0,$ and $\Omega_c^0$ 
particles and a neutrino branch with the  $\pi^\pm$, K$^{\pm,0}$, n, 
D$^{\pm,0}$, and D$_S^\pm$ particles, as follows clearly from 
the decays of the particles. The masses of the particles of the 
$\gamma$-branch are integer multiples of the mass of the $\pi^0$ meson or 
are proportional to an integer quantum number n, on 
the average within 0.66\%. The masses of the particles of the $\nu$-branch are integer 
multiples of the mass of the $\pi^\pm$ mesons times a factor 0.85 $\pm$ 
0.02. Nambu [6] has first suggested that the masses of the elementary 
particles might show some regularity ``if the masses were measured in a 
unit of the order of the $\pi$-meson mass". Barut [7] has shown  
that the masses of the e, $\mu,\,\tau$ leptons can be described with very 
good accuracy by the fourth power of a principal integer quantum number n. 
This rule has been extended by Gsponer and Hurni [8] to the masses of the 
quarks.

  We have also shown that the masses of the 
particles of the $\gamma$-branch can be explained by standing 
electromagnetic waves in a cubic cavity. The standing waves come with a 
continuum of frequencies which must be in the particles after a high 
energy collision or according to Fourier analysis. The 
particles of the $\nu$-branch consist of cubic neutrino lattices. The 
energy in the $\nu$-branch particles is the energy of the lattice 
oscillations plus the energy in the rest masses of the neutrinos.

\section{The spin of the particles of the $\gamma$-branch}

   We will now 
show that the spin of the particles can be explained by the sum of the angular 
momentum vectors and the spin vectors of the waves and neutrinos and 
electric charges in the mesons and baryons. It is striking that the 
particles which, according to the standing wave model, 
consist of a single oscillation mode do not have spin, as the $\pi^0, 
\pi^\pm$ and $\eta$ mesons do, see Tables 1 and 2. The modes in these tables are 
improved versions of the modes in Tables 1 and 4 in [5], in which the consequences of 
electrical charge on the structure of the particles were not considered. 
 It is also striking that particles whose mass is approximately 
twice the mass of a smaller particle have spin 1/2 as is the 
case with the $\Lambda$ baryon, m($\Lambda$) $\approx$ 2m($\eta$), and with the 
nucleon m(n) $\approx$ 2m(K$^\pm$) $\approx$ 2m(K$^0$). The $\Xi^0_c$ baryon which is a doublet of 
one mode has also spin 1/2. Composite particles 
which consist of a doublet of one mode plus one or two other single modes have spin 1/2, 
as the $\Sigma^0$, $\Xi^0$\, and $\Lambda_c^+$, $\Sigma_c^0$, $\Omega_c^0$ 
baryons do. The only particle which seems to be a triplet of a single mode, the $\Omega^-$ 
baryon, has spin 3/2. It appears that the relation between the spin and 
the oscillation modes of the particles is straightforward.

\begin{table}\caption{ The spin and oscillation modes of the 
$\gamma$-branch particles. The $\cdot$ marks coupled oscillations.}
\hspace{3cm}\begin{tabular}{lcll}\\
\hline\hline\\
particle & mode & spin\\
[0.2ex]\hline\\
$\pi^0$ & (1.1)$\pi^0$ & 0\\
$\eta$ & (2.2)$\pi^0$ & 0\\
$\Lambda$ & 2$\cdot$(2.2)$\pi^0$ & 1/2\\
$\Sigma^0$ & 2$\cdot$(2.2)$\pi^0$ + $\pi^0$ & 1/2\\
$\Sigma^\pm$ & 2$\cdot$(2.2)$\pi^0$ + $\pi^\pm$ & 1/2\\
$\Xi^0$ & 2$\cdot$(2.2)$\pi^0$ + 2$\pi^0$ & 1/2\\
$\Xi^-$ & 2$\cdot$(2.2)$\pi^0$ + $\pi^0 + \pi^-$ & 1/2\\
$\Omega^-$ & 3$\cdot$(2.2)$\pi^0$ + $\pi^-$ & 3/2\\
$\Lambda^+_c$ & 2$\cdot$(2.2)$\pi^0$ + 2K$^0$ + $\pi^+$ & 1/2\\
$\Sigma^0_c$ & $\Lambda^+_c$ + $\pi^-$ & 1/2\\
$\Xi^0_c$ & 2$\cdot$(3.3)$\pi^0$ & 1/2\\
$\Omega^0_c$ & 2$\cdot$(3.3)$\pi^0$ + 2$\pi^0$ & 1/2\\
[0.2cm]\hline\hline\\
\end{tabular}\\
\end{table}

\begin{table}\caption{ The spin and oscillation modes
 of the $\nu$-branch particles. (2.2)$\pi^\pm_n$ = 340 MeV is the second mode 
 of the neutrino lattice oscillations plus the rest masses of the 
 neutrinos in the lattice, as discussed in [5], section 6. }
\hspace{1cm}\begin{tabular}{lcll}\\
\hline\hline\\
particle & mode & spin\\
[0.2ex]\hline\\
$\pi^\pm$ & (1.1)$\pi^\pm$ & 0\\
K$^\pm$ & (2.2)$\pi^\pm_n$ + $\pi^0$ & 0\\
K$^0$ & (2.2)$\pi^\pm_n$ + $\pi^\mp$ & 0\\
n & 2$\cdot$(2.2)$\pi^\pm_n$ + 2$\pi^\mp$ & 1/2\\
D$^{\pm}$ & 2[(2.2)$\pi^\pm_n$ + $\pi^\mp$] + (2.2)$\pi^\pm_n$ +  
$\pi^0$ + (2.2)$\pi^\mp_n$ + $\pi^\pm$ & 0\\
D$^0$ & 2[(2.2)$\pi^\pm_n$ + $\pi^\mp$] + (2.2)$\pi^+_n$ + (2.2)$\pi^-_n$ + 
2$\pi^0$ & 0\\
D${^\pm_S}$ & 4(2.2)$\pi^\pm_n$ + 3$\pi^\mp$ + 2$\pi^0$ & 0\\
[0.2cm]\hline\hline\\
\end{tabular}\\
\end{table}

  In the standing wave model  the $\pi^0$ and $\eta$ mesons consist 
of N = 2.85$\cdot10^9$ standing electromagnetic waves, each with its own frequency. The 
oscillations in the photon lattice are longitudinal. All 
standing longitudinal waves of frequency $\nu_i$ in the $\pi^0$ and 
$\eta$   mesons do not have angular momentum or 
 $\sum_{i}j(\nu_i)$ = 0, with the index running from 1 to N. 
 Longitudinal lattice oscillations 
 cannot cause an intrinsic angular momentum because for longitudinal 
 oscillations $\vec{r}\,\times\,\vec{p}$ = 0.

   Each of the standing electromagnetic waves in 
the $\pi^0$ and $\eta$ mesons may, on the other hand, have spin s = 1 of its 
own, because circularly polarized electromagnetic waves have an angular 
momentum as was first suggested by Poynting [9] and verified by, 
among others, Allen [10]. The creation of the $\pi^0$ meson in the reaction 
$\gamma \,+\, p \rightarrow \pi^0$ + p and conservation of angular 
momentum dictate that the sum of the angular momentum 
vectors of the N electromagnetic waves in the $\pi^0$ meson is 
zero, $\sum_i\,j(s_i)$ = 0.  Either the 
sum of the spin vectors of the electromagnetic 
waves in the $\pi^0$ meson is zero, or each electromagnetic wave in the 
$\pi^0$ meson has zero spin. That 
would mean that they are linearly 
polarized. Linearly polarized electromagnetic waves are not expected to have 
angular momentum. That this is actually so was proven by Allen [10] who 
specifically states that ``a wave linearly polarized to the [receiving] 
dipole" and ``a wave linearly polarized perpendicular to the dipole" does ``in 
neither instance ... exert a torque on the dipole", which means that these 
waves do not 
have angular momentum, whereas ``the dipole will experience a torque" when 
the ``wave is circularly polarized" and consequently the wave has 
angular momentum. Since the standing longitudinal photon 
oscillations in the $\pi^0$ and $\eta$ mesons do not have 
angular momentum and since the sum of the spin vectors $s_i$ of the 
electromagnetic  waves must be zero, the intrinsic angular 
momentum of the $\pi^0$ and $\eta$ mesons is zero, or
\begin{equation} \sum_{i}\,(j(\nu_i) + j(s_i)) = 0\quad (1\,\le\,i\le N)\,.\end{equation}
In the standing wave model the $\pi^0$ and $\eta$ mesons 
do not have intrinsic angular momentum or 
spin, as it must be.

  We now consider particles such as the $\Lambda$ baryon 
which consist of superpositions of two perpendicular 
standing waves of equal frequencies and amplitudes shifted 
in phase by $\pi$/2 at each of the N points of the 
lattice [5]. The oscillations in the particles are then coupled what we have 
marked in Tables 1,2 by the $\cdot$ sign. The particles then contain 
N circular waves, each with its own frequency and each 
having an angular momentum of $\hbar$/2 as we will see.

  The superposition of two perpendicular linearly polarized traveling 
waves of equal amplitudes and frequencies shifted in phase by $\pi$/2 
leads to a circular wave with 
the constant angular momentum j = $\hbar$. The total 
energy of such a wave is the sum of the potential and the 
kinetic energy. If the motion is circular the 
kinetic energy is always equal to the potential energy. From 

\begin {equation} E_{pot} + E_{kin} = E_{tot} = \hbar\omega\,,\end {equation}
\noindent
follows \begin{equation} E_{tot} = 2E_{kin} = 
2\frac{\Theta\omega^2}{2}\, = \hbar\omega 
,\end{equation}
\noindent
with the moment of inertia $\Theta$. It follows that the angular momentum 
j is
 
\begin{equation} j = \Theta\omega = \hbar\,.\end{equation}
\noindent
   This applies for one single circular 
oscillation and corresponds to spin s 
= 1, or to a circularly polarized electromagnetic wave.

  We now add to one monochromatic standing 
oscillation with frequency $\omega$ a  perpendicular second standing oscillation 
with the same frequency shifted in phase by 
$\pi$/2, having the same amplitude, as we have done before in [11]. In 
other words we consider the oscillations

\begin{equation} x(t) = exp[i\omega t] + exp[-\,i(\omega t + 
\pi)]\,,\end{equation}

\begin{equation} y(t) = exp[i(\omega t + \pi/2)] + exp[-\,i(\omega t + 
                 3\pi/2)]\,\,.
\end{equation}
\noindent
If we replace \emph{i} in Eqs.\,(5,6) by $-$\,\emph{i} we have a circular wave turning in 
opposite direction.
The energy of the superposition of the two waves is the sum of the 
energies of both individual waves, so according to Eq.(2) we have

\begin{equation} 4E_{kin} = 4\Theta\omega^2/2 = E_{tot} = 
\hbar\omega\,,\end{equation}
\noindent
from which follows that the standing circular wave has an angular momentum

\begin{equation} j = \Theta\omega = \hbar/2\,. \end{equation}
\noindent
The superposition of two perpendicular monochromatic standing waves of equal amplitudes 
and frequencies shifted in phase by $\pi$/2 satisfies 
the necessary condition for spin s = 1/2 that the angular momentum is 
j = $\hbar$/2.

   The standing wave model of the mesons and baryons treats the $\Lambda$ 
baryon, which has spin s = 1/2 and a mass m($\Lambda$) = 
1.0190\,$\cdot$\,2m($\eta$), as the superposition of two  particles of the 
same type 
with N standing electromagnetic waves. The waves are circular because 
they are the superposition of two standing waves with the same 
frequency and amplitude shifted in phase by $\pi$/2.  The angular 
momentum  vectors  around a center axis of all circular waves in the lattice cancel,   
except for the wave at the center of the crystal, because for each 
oscillation with frequency $\omega$ 
there is at its mirror position a wave with the frequency $-\,\omega$, 
which consequently has a negative angular momentum since j = mr$^2\omega$. Oscillations with 
negative frequencies are permitted solutions in cubic isotropic lattices.

   The frequency distribution of the axial longitudinal oscillations of a 
cubic lattice has to be corrected for the limitation of the group 
velocity to the value of the velocity of light. The oscillation frequencies 
are then given, according to [5], by
\begin{equation} \nu = \nu_0\phi = c\,\phi/2\pi\emph{a}\,, \end{equation}
\noindent
with $\phi = 2\pi\emph{a}/\lambda$, the lattice constant $\emph{a} = 
10^{-16}$\,cm, the 
wavelength $\lambda$, and the velocity of light c. Since $\phi$ goes from $-\,\pi$ to $\pi$ there is for 
each positive frequency a negative frequency of the same absolute value 
at $-\,\phi$. Consequently the angular momentum vectors of all circular 
waves in the lattice cancel, but for the wave at the center of the 
lattice, whose frequency seems to be zero according to Eq.(9). This is a misleading result of 
the assumption of an infinite extension of the crystal introduced by the 
periodic boundary conditions used in the classic theory of lattice 
oscillations. The frequency at $\phi = 0$ is $\nu(0) = c/2d$, where d is 
the distance between two opposite sides of the crystal, d $\approx$ 
10$^3\,\emph{a}$. As for all circular oscillations in the lattice the wave at the 
center is also the superposition of two perpendicular standing waves and 
has an angular momentum of $\hbar$/2 according to Eq.(8). This is 
the only one of the waves whose angular momentum is not canceled. The net 
angular momentum of the N lattice oscillations which are the superpositions 
of two perpendicular standing waves is therefore $\hbar$/2.  Since the 
circular standing waves in the $\Lambda$ baryon lattice are the 
only possible contribution to an angular momentum the 
intrinsic angular momentum of the $\Lambda$ baryon is $\hbar$/2 or
\begin{equation} j(\Lambda) = \sum_i\,j(\nu_i) =  \hbar/2\,.\end{equation}             
 We have thus explained that the 
$\Lambda$ and likewise the $\Xi_c^0$ baryon satisfy the necessary condition 
that j = $\hbar$/2 for s = 1/2. The intrinsic angular momentum of the 
$\Lambda$ baryon is the consequence of the superposition of two 
perpendicular standing waves of the same frequency shifted in phase by 
$\pi$/2.

   The other particles of the $\gamma$-branch, the 
$\Sigma^0$, $\Xi^0$, $\Lambda^+_c$, $\Sigma^0_c$ and $\Omega^0_c$ baryons are 
 composites of a baryon with spin 1/2 plus one or two $\pi$ 
mesons which do not have spin. Consequently the spin of these particles is 
1/2. The spin of all particles of the $\gamma$-branch, exempting the 
spin of the $\Omega^-$ baryon, has thus been explained.    

\section{The spin of the particles of the $\nu$-branch}

The characteristic particles of the neutrino-branch are the $\pi^\pm$ mesons 
which have zero spin. At first glance it seems to be odd that the $\pi^\pm$ 
mesons do not have spin, because it seems that the $\pi^\pm$ mesons should have 
spin 1/2 from the spin of the charges e$^\pm$ in $\pi^\pm$. However that is not the case. The 
solution of this puzzle is in the composition of the $\pi^\pm$ mesons 
which are, according to the standing wave model, made of a lattice of 
neutrinos and antineutrinos (Fig.\,1), each having spin 1/2, whereas the $\pi^0$ 
meson is 
made of standing electromagnetic waves which are linearly polarized and 
do not have spin.

	\begin{figure}[h]
	\hspace{3cm}
	\includegraphics{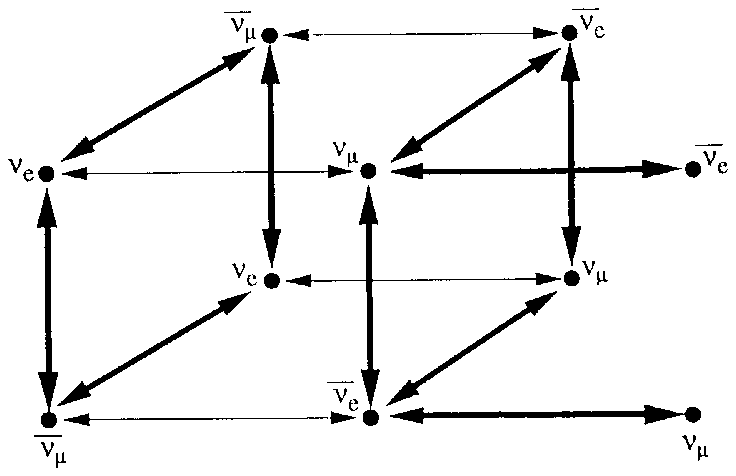}
	\vspace{-0.2cm}
	\begin{quote}
Fig.\,1: The neutrino lattice of the $\pi^\pm$ mesons. After [5].
	\end{quote}
	\end{figure}

  In a cubic lattice of N = 2.85$\cdot$$10^9$ neutrinos and antineutrinos the 
spin of nearly all neutrinos must cancel because conservation 
of angular momentum during the creation of the particle requires that the 
total angular momentum around a central axis is only a small integer or half-integer number. In 
fact the spin vectors of all but the neutrino in the center of 
the lattice cancel.  In order for this to be so 
the direction of the spin of any particular neutrino in the lattice 
has to be opposite to the direction of the spin of the antineutrino at its 
mirror position. Neutrinos and antineutrinos of the same type differ in the 
lattice only by the direction of their spin. As can be seen on Fig.\,1 
each neutrino has at its mirror position an antineutrino.  The only angular 
momentum remaining from the spin of the neutrinos of the lattice is the 
angular momentum of the neutrino at the center of the lattice. 
Consequently the electrically neutral neutrino lattice consisting of 
N neutrinos, each with a spin $n_i$, has an intrinsic 
angular momentum j = $\sum_i\,j(n_i)$ = $\hbar$/2.

  The standing longitudinal oscillations of the neutrinos in the lattice 
of the $\pi^\pm$ mesons do not 
cause an angular momentum, $\sum_i\,j(\nu_i)$ = 0. But electrons or positrons added to the neutral 
neutrino lattice have spin 1/2. If the spin of the electron or positron 
added to the neutrino lattice is opposite to the spin of the neutrino in 
the center of the lattice then the net spin of the $\pi^+$ or $\pi^-$ mesons is 
zero, or
 \begin{equation} j(\pi^\pm) = \sum_i\,j(n_i) + j(e^\pm) = 0\,\quad(1\leq 
 i \leq N)\,.
\end{equation}
It is important for the understanding of the structure of 
the $\pi^\pm$ mesons to realize that s($\pi^\pm$) = 0 can only be 
explained if the $\pi^\pm$ mesons consist of a \emph{neutrino lattice} 
to which an electron or positron is added whose spin is opposite to the net 
spin of the neutrino lattice. Spin 1/2 of the electric charges can only be 
canceled by something that has also spin 1/2, and the only conventional 
choice for that is the neutrino.

   The spin of the K$^{\pm}$ mesons is zero. With the spin of the K$^\pm$ mesons we 
encounter the same oddity we have just observed with the spin of the 
$\pi^\pm$ mesons, namely we have a particle which carries an electrical charge with spin 
1/2, and nevertheless the particle does not have spin. The explanation 
of s(K$^\pm$) = 0 follows the same lines as the explanation of 
the spin of the $\pi^\pm$ mesons. In the standing wave model the K$^\pm$ 
mesons are described by the state (2.2)$\pi^\pm_n$ + $\pi^0$, that means by the (2.2) 
mode of the charged neutrino lattice oscillations plus 
a $\pi^0$ meson. The (2.2)$\pi^\pm_n$ state of the neutrino lattice is the 
sum of the energy of the (2.2) oscillation of the lattice plus the sum of the energies 
of the rest masses of the neutrinos and contains 340 Mev. The (2.2) 
mode of the longitudinal oscillations of a neutral neutrino lattice  
 does not have a net intrinsic angular momentum. But the spin 
of the neutrinos contributes an angular momentum $\hbar$/2, which originates 
from the neutrino in the center of the lattice, just as it 
is with the neutrino lattice in 
the $\pi^\pm$ mesons, so $\sum_i\,{ j(n_i)} = \hbar/2$. Adding 
an electric charge with a spin opposite to the net intrinsic angular 
momentum of the neutrino lattice creates the charged (2.2)$\pi^\pm_n$ mode 
which has zero spin,

\begin{equation}j((2.2)\pi^\pm_n) = \sum_i\,{ j(n_i)} + j(e^\pm) = 
0\,.
\end{equation}
 As discussed in [5] it is 
necessary to add a $\pi^0$ meson to the (2.2)$\pi^\pm_n$ mode of the 
$\pi^\pm$   meson in 
order to obtain the correct mass and the correct decays of the K$^\pm$ mesons. Since the $\pi^0$ 
does not have spin the addition of the $\pi^0$ meson does not add to the 
intrinsic angular momentum of the $K^\pm$ mesons. So, according to 
Eq.(12), s(K$^\pm$) = 0 in agreement with the facts.

  The explanation of s = 0 of the K$^0$ mesons described by 
the state (2.2)$\pi^\pm_n$ + $\pi^\mp$ follows similar lines. The 
longitudinal oscillations of the (2.2)$\pi^\pm_n$ mode as 
well as of the basic $\pi^\mp$ mode do not create an angular momentum, 
$\sum_i\,j(\nu_i)$ = 0. The first 
higher mode of the $\pi^\pm$ mesons, the (2.2)$\pi^\pm_n$ state, and the 
basic $\pi^\mp$ mode each have N neutrinos, so the 
number of neutrinos in the sum of both states, 
the K$^0$ meson, is 2N. Since the size of the lattice of the K$^\pm$ mesons and 
the K$^0$ mesons is the same it follows that two neutrinos are at each lattice 
point of the K$^0$ meson. We assume that Pauli's exclusion principle applies 
for neutrinos as well. Consequently each neutrino at each lattice point must 
share its location with an antineutrino. That means that the contribution 
of the spin of all neutrinos to the intrinsic angular momentum of the 
K$^0$ meson is zero or $\sum_i\,j(2n_i)$ = 0.  The 
sum of the spin vectors of the two opposite charges in the K$^0$ meson, 
i.e. in (2.2)$\pi^\pm$ + $\pi^\mp$, is also 
zero. Since neither the neutrino oscillations nor the spin of the neutrinos 
nor the  electric charges contribute an angular momentum or since 
\begin{equation} j(K^0) = \sum_i(\,j(\nu_i) + j(2n_i)) + j(e^+ + e^-) = 0\,,
\end{equation}
 the intrinsic angular 
momentum of the K$^0$ meson is zero, or s(K$^0$) = 0, as it 
must be. In simple terms, since the structure of K$^0$ is 
(2.2)$\pi^\pm_n$ + $\pi^\mp$, the spin of K$^0$ is the sum of the spin of  
(2.2)$\pi^\pm_n$ and $\pi^\mp$ in K$^0$, both of which do not have spin. It does not 
seem possible to arrive at s = 0 for the K$^0$ meson if the particle 
does not contain the 2N neutrinos required by the (2.2)$\pi^\pm_n$ + $\pi^\mp$ 
state which we have suggested in section 6 of [5].  

   In the standing wave model the neutron, which has spin 1/2 and a mass 
$\approx$ 2m(K$^\pm$) or 2m(K$^0$), is either the 
superposition of a K$^+$ and a K$^-$ meson or of two K$^0$ mesons. In the case 
of the neutron 
one must wonder how it comes about that a particle which seems to be the 
superposition of two particles without spin ends up with spin 1/2.  
The intrinsic angular 
momentum of the sum of K$^+$ and K$^-$ comes in part from the superposition of two perpendicular standing 
longitudinal neutrino lattice oscillations. 
Similar to the case of the $\Lambda$ baryon the sum of the angular momentum 
vectors of the circular oscillations of all neutrinos in the lattice 
reduces to the angular momentum of the circular wave at the center of the 
lattice. Following Eq.(8) the angular momentum of the circular wave at the center 
or the net angular momentum of the entire neutrino lattice oscillations 
is $\sum_i\,j(\nu_i)$ = $\hbar$/2. The intrinsic angular momentum of the 
neutron consists 
consequently in part of the angular momentum of the circular neutrino 
lattice oscillations whose net sum is $\hbar$/2.

  The superposition of two K$^\pm$ mesons also means that the 
lattice contains 2N neutrinos because K$^+$ as well as K$^-$ each contains 
N neutrinos. 
Assuming that Pauli's exclusion principle applies, each 
neutrino in the lattice must share its place with a neutrino of 
opposite spin. That means that the spin vectors of all neutrinos of the 
lattice cancel, $\sum_i\,j(2n_i)$ = 0.  Neither the neutrinos nor the 
spin of the charges of the sum of the K$^+$ and K$^-$ mesons contribute to the net 
intrinsic angular momentum. But, and that is crucial, the superposition of 
the oscillations in 
the two $\pi^0$ mesons which are part of the (2.2)$\pi^\pm_n$ + $\pi^0$ 
structure of K$^\pm$ adds 
an angular momentum $\sum_i\,j(2\pi^0)$ = $\hbar$/2 to the net angular 
momentum of the sum of K$^+$ + K$^-$, because the electromagnetic 
waves in the $\pi^0$ mesons are 
superposed at right angles and shifted in phase by $\pi$/2 just as the 
neutrino lattice oscillations.  Consequently the sum of the intrinsic angular momentum vectors  of the 
superposition of a K$^+$ and a K$^-$ meson is 
\begin{equation} j(K^+ + K^-) = \sum_i( j(\nu_i) + j(2n_i)) + j(e^+ + e^-) + 
j(2\pi^0) = 0\,\, \mathrm{or}\,\, \hbar\,, \end{equation}
 which is 
incompatible with the experimental facts. We conclude that the neutron 
\emph{cannot} be the superposition of a K$^+$ and a K$^-$ meson.

   On the other hand the neutron can well be the superposition of two 
K$^0$ mesons or of a K$^0$ and a $\mathrm{\bar{K}}^0$ meson. A significant change 
in the lattice occurs when two K$^0$ 
mesons are superposed. Since each K$^0$ meson contains 2N neutrinos, as we 
discussed before in context with the spin of K$^0$, the number of 
neutrinos in two superposed K$^0$ lattices is 4N. Since the size of the 
lattice of the proton as well of the neutron is the same as the size of 
K$^0$, (the measured 
r$_p$ is within the experimental error the same as r$_\pi$), each lattice 
point now contains four neutrinos, a muon neutrino and an anti-muon 
neutrino as well as an 
electron neutrino and an anti-electron neutrino. The quartet of neutrinos 
oscillates just like individual neutrinos do because we have found in [5] 
that the ratios of the sum of the oscillation frequencies are independent 
of the mass as well as of the interaction constant between the lattice 
points. In the neutrino quartets the spin of the 
neutrinos cancels, $\sum_i\,j(4n_i)$ = 0. The superposition of the 
neutrino  oscillations,  that means of circular oscillations of frequency 
$\nu_i$,    contribute the 
angular momentum of the center circular wave, so 
$\sum_i\,j(\nu_i) = \hbar/2$. 
The spin and charge of the four electrical charges hidden in the two 
K$^0$ mesons cancel. There is no $\pi^0$ component in K$^0$ and 
consequently no contribution to the intrinsic angular momentum. 
It follows that the intrinsic angular momentum of a neutron created by the 
superposition of two K$^0$ mesons comes from the circular neutrino lattice 
oscillations only and is 
\begin{equation} j(n) = \sum_i(\,j(\nu_i) + j(4n_i)) + j(4e^\pm) = \hbar/2\,,
\end{equation}
as it must be. In simple terms, the spin of the neutron originates from 
the superposition of two perpendicular standing neutrino lattice 
oscillation  with the same frequencies shifted in phase by $\pi$/2, which 
produces the angular momentum $\hbar$/2. 

   The spin of the proton is 1/2 and is unambiguosly defined by the decay 
of the neutron n $\rightarrow$ p + e$^-$ + $\bar{\nu}_e$. The 
remaining particles of the neutrino branch, the D$^{\pm,0}$ and 
D$_S^\pm$ mesons both having zero spin, 
are superpositions of a proton and a neutron of opposite spin, or of 
their antiparticles, or of two neutrons of opposite spin in D$^0$.  
The spin of D$^\pm$ and D$^0$ does therefore not pose a new 
problem. We note in passing that in D$^0$ a $\pi^0$ meson replaces a 
$\pi^\pm$ meson in D$^\pm$, and consequently the mass difference 
m(D$^\pm$)\,$-$\,m(D$^0$) = 4.78 MeV is very similar to m($\pi^\pm$)\,$-$\,m($\pi^0$) = 4.593 
MeV, comparable to the case m(K$^0$)\,$-$\,m(K$^\pm$) in [5]. Replacing a $\pi^\pm$ 
meson by a $\pi^0$ meson does not change the spin. In D$^\pm_S$ only 
a $\pi^0$ meson is added to D$^\pm$ and 
therefore its spin is the same as that of D$^\pm$.

   The $\Sigma^+$, the $\Sigma^-$ and the $\Xi^-$ baryons are afflicted by 
the same problem we have encountered with the $\pi^\pm$ and K$^\pm$ mesons, 
namely their spin does not seem to add up. When electric charge, 
either an electron or positron with spin 1/2, is added to the $\Sigma^0$ 
or $\Xi^0$ baryons, both of which have spin 1/2, the resulting $\Sigma^\pm$ 
and $\Xi^-$ baryons should not have the same spin s = 1/2 as the 
$\Sigma^0$ and $\Xi^0$ have. This puzzle can be solved when the charge 
added to the $\Sigma^0$ or $\Xi^0$ is not added as an electron or positron 
but rather as $\pi^+$ or $\pi^-$ mesons which do not have spin. 
Considering the structure of the $\Sigma^0$ baryon we have found in [5], 
which is 2$\cdot$(2.2)$\pi^0$ + $\pi^0$, we can replace the single 
$\pi^0$ meson in $\Sigma^0$ with a $\pi^+$ or $\pi^-$ meson without changing 
the mass of $\Sigma^0$ significantly and without changing the spin. The 
same applies correspondingly to 
$\Xi^0$. The presence of either $\pi^+$ or $\pi^-$ in $\Sigma^\pm$ becomes 
quite clear from the decay of these particles. $\Sigma^+$ decays into 
p + $\pi^0$ (51.57\%) and n + $\pi^+$ (48.31\%), both make up 99.88\% of 
the $\Sigma^+$ decays. $\Sigma^-$ decays into $n\,+\,
\pi^-$ (99.848\%), whereas $\Sigma^0$ decays into 
$\Lambda\,+\,\gamma$ (100\%). $\Xi^-$ decays into $\Lambda\,+\,\pi^-$ (99.887\%) 
whereas $\Xi^0$ decays into $\Lambda\,+\,\pi^0$ (99.54\%). In 
particular in the latter case it seems to be obvious that in the $\Xi^-$ 
baryon a 
$\pi^-$ meson replaces a $\pi^0$ meson in $\Xi^0$. If this is the 
case, then there is no 
problem with the spin of $\Xi^-$ being the same as the spin of $\Xi^0$. 
The same applies to the spin of the $\Sigma^\pm$ baryons.

   An explanation of the spin of the mesons and baryons can only be valid if the same explanation also applies to the antiparticles of these particles whose spin is the $\emph{same}$ as that of the ordinary particles. In the standing wave model the antiparticles of the $\gamma$-branch consist of the same photons as in the ordinary $\gamma$-branch particles, other than that the sign of the frequency of each of the standing electromagnetic waves is reversed. If a particle, such as the $\Lambda$ baryon, consists of superpositions of standing perpendicular waves shifted in phase by $\pi$/2, all waves in the cubic lattice of the particle are circular, and the angular momentum vectors of all waves in the lattice cancel, but for the wave at the center, regardless whether the frequencies are positive or negative. The remaining wave at the center has an angular momentum of $\hbar$/2, regardless whether the frequencies are positive or negative. Reversing the sign of the frequency means only a phase shift of $\pi$. Consequently the intrinsic angular momentum of the antiparticles of the $\gamma$-branch particles have the same spin s = 1/2 as the ordinary particles. Of the particles of the $\nu$-branch only the neutron has spin s = 1/2. As shown above the spin of the neutron is a consequence of circular neutrino lattice oscillations. If we replace the positive frequencies of the neutrino lattice oscillations in the neutron with negative oscillations in the antineutrino then only the phase of the oscillations is shifted by $\pi$, but the net angular momentum of all circular oscillations is preserved. Consequently the spin of the antineutron is the same as the spin of the neutron, as it must be.        

\bigskip

\section*{Conclusions}

   The intrinsic angular momentum of the $\gamma$-branch of the so-called stable 
elementary particles can be explained with the sum of the angular momentum 
vectors of the electromagnetic waves or photon lattice oscillations, plus the sum of 
the spin vectors of the electromagnetic waves, plus the 
sum of the spin vectors of the electric charges if any are in the particles 
of the $\gamma$-branch. Correspondingly, the intrinsic 
angular momentum of the particles of the neutrino branch is the sum of the 
angular momentum vectors of the neutrino lattice oscillations, plus the sum of the 
spin vectors of the neutrinos, plus the sum of the spin vectors of the 
electric charges which the particles carry.  

   The most simple particles, the $\pi^0$ and $\eta$ mesons, do not have spin 
because neither the longitudinal oscillations of the photon lattice nor the 
linearly polarized electromagnetic waves can contribute to an intrinsic angular 
momentum. The $\pi^\pm$ and K$^\pm$ mesons are quite different because the electric 
charges in either particle bring with them spin 1/2. Since s($\pi^\pm$) 
and s(K$^\pm$) = 0 the spin of e$^\pm$ in $\pi^\pm$ and K$^\pm$ must be 
canceled by something that has likewise spin 1/2. We have shown that the 
sum of the spin vectors of the very many neutrinos which in our model 
are in $\pi^\pm$ and K$^\pm$ reduces to the spin of the neutrino at the 
center of the lattice. If the spin of the electric charge and of the 
 neutrino at the center of the lattice are of opposite direction 
 then s($\pi^\pm$) and s(K$^\pm$) = 0, as it must be. 
The spin of K$^0$ is likewise zero. There can be no contribution to the 
spin of K$^0$ from the two opposite charges which are in K$^0$ according to 
our model. There is also no contribution from the spin of the neutrinos 
because at each lattice point the spin of each neutrino is canceled by the 
spin of its antineutrino. Neither the longitudinal lattice oscillations, 
nor the spin vectors of the neutrinos, nor the two electrical charges 
contribute to an angular momentum of K$^0$, so s(K$^0$) = 0.

   The spin s = 1/2 of the $\Lambda$ baryon and the neutron, which have a 
mass m($\Lambda$) $\approx$ 2m($\eta$) and m(n) $\approx$ 2m(K$^0)$, can be 
explained with the sum of the angular momentum vectors of the 
superpositions of two 
perpendicular standing waves of the same frequencies shifted in 
phase by $\pi$/2, as they are in 
$\Lambda$ and n in the standing wave model. In either case an angular 
momentum $\hbar$/2 remains at the center of the lattice from the multitude 
of standing circular waves in these particles. There can be no other 
contributions to the intrinsic angular momentum of these particles and 
hence s($\Lambda$) and s(n) = 1/2. The other baryons are composites of 
$\Lambda$ and their spin does not pose a new problem.

   Our explanation of the intrinsic angular momentum confirms the 
validity of the structure of the particles which we considered. A 
convincing example of 
this correlation is offered by the explanation of the puzzling absence 
of spin in the $\pi^\pm$ mesons in spite of the electric charge these 
particles carry. 
The standing wave model assumes that the $\pi^\pm$ mesons have a 
cubic neutrino lattice. The net spin of the neutrinos in the 
lattice cancels the spin of the 
electric charges in $\pi^\pm$. The spin, the mass and the 
decays of the $\pi^\pm$ mesons require a neutrino lattice for 
the $\pi^\pm$ mesons. Spin 1/2 of the $\Lambda$ baryon, 
and consequently of the other baryons of the $\gamma$-branch, and spin 1/2 
of the neutron originates from the superposition of two perpendicular 
standing oscillations in the particles, as the standing wave model 
postulates; and as is indicated by the mass ratios 
m($\Lambda$)/m($\eta$) $\approx$ 2 and m(n)/m(K$^0$) $\approx$ 2. The spin 
of the stable mesons and baryons can be explained with the standing wave model 
without an additional assumption.

\bigskip

\noindent
\textbf{REFERENCES}

\bigskip

\noindent
[1] Uhlenbeck,G.E. and Goudsmit,S. Naturwiss. {\bfseries13},953 (1925).\\
\smallskip
\noindent
[2] Biedenharn,L.E. and Louck,J.D. \emph{Angular Momentum in Quantum}\\ 
\indent\emph{Physics},\,\, Addison Wesley, (1981).\\
\smallskip
\noindent
[3] Jaffe,R.L. arXiv: hep-ph/0008038 (2000).\\
\smallskip
\noindent
[4] G\"ockeler,M. et al. arXiv: hep-lat/0208017 (2002).\\
\smallskip
\noindent
[5] Koschmieder,E.L. arXiv: phys/0211100 (2002).\\
\smallskip
\noindent
[6] Nambu,Y. Progr.Th.Phys. {\bfseries7},595 (1952).\\
\smallskip
\noindent
[7] Barut,A.O. Phys.Rev.Lett. {\bfseries42},1251 (1979).\\
\smallskip
\noindent
[8] Gsponer,A. and Hurni,J-P. Hadronic J. {\bfseries19},367 (1996).\\
\smallskip
\noindent
[9] Poynting,J.H. Proc.Roy.Soc. {\bfseries A 82},560 (1909).\\
\smallskip
\noindent
[10] Allen,J.P. Am.J.Phys. {\bfseries34},1185 (1966).\\
\smallskip
\noindent
[11] Koschmieder,E.L. Nuovo Cim. {\bfseries99},555 (1988).

\end{document}